\documentclass[12pt,notitlepage]{article}
\usepackage[T1]{fontenc}
\usepackage[utf8]{inputenc}
\usepackage{amsmath,amssymb,amsthm}
\usepackage{upgreek}
\usepackage{mathpazo,setspace,xcolor}
\usepackage{booktabs,enumitem,microtype,needspace}
\usepackage[round,authoryear]{natbib}
\usepackage[top=0.9in,bottom=0.9in,left=0.9in,right=0.9in]{geometry}
\usepackage{footmisc}
\usepackage[colorlinks=true,citecolor=blue,linkcolor=blue,urlcolor=blue]{hyperref}
\usepackage{bookmark}

\newtheorem{proposition}{Proposition}
\newtheorem{lemma}{Lemma}
\newtheorem{corollary}{Corollary}
\theoremstyle{definition}
\newtheorem{definition}{Definition}
\newtheorem{example}{Example}
\newtheorem{remark}{Remark}
\numberwithin{equation}{section}
\setlist[enumerate]{itemsep=3pt,topsep=5pt}

\newcommand{\HIR}{\mathrm{HIR}}
\newcommand{\R}{\mathbb R}

\DeclareMathOperator*{\argmax}{arg\,max}

\begin{document}
\title{Stable matching under regulation}
\author{Chao Huang\thanks{Institute for Social and Economic Research, Nanjing Audit University. Email: huangchao916@163.com.}}
\date{}
\maketitle

\begin{abstract}
In a market in which any coalition can reach agreements, a stable outcome need not exist even when agents have substitutable preferences. We show how regulation can restore stability. If a regulator participates in every agreement, a setwise stable outcome exists when each ordinary agent's preference satisfies a condition weaker than substitutability, while the regulator's preference remains largely unrestricted. An alternative existence result allows ordinary agents to have arbitrary preferences while requiring the regulator to support each agent's profitable contract deletions. Our results apply to matching markets involving regulated activities.
\end{abstract}


\section{Introduction}\label{sec:introduction}

The existence of stable matchings depends on both preferences and market structure. Stable outcomes exist in two-sided matching markets under substitutable preferences (e.g., \citealp{KC82,RS90}) and in multilateral matching markets under complementary preferences (e.g., \citealp{RY20,RY25}).

Substitutability alone does not guarantee existence when any coalition can reach agreements. The stable roommates problem is a special case of this setting and already exhibits nonexistence with unit-demand agents (e.g., \citealp{GS62,I85,T91}). We show how regulation can restore the existence of setwise stable outcomes. If a regulator participates in every agreement, a setwise stable outcome exists when each ordinary agent's preference satisfies \textbf{hereditary individual rationality} (HIR). Individual rationality rules out profitable deletions of existing contracts. HIR requires every subset of an individually rational bundle to remain individually rational.

Our results apply to finite-contract markets with indifferences (e.g., \citealp{I94,S99,EE08,EE17,RY20}) and to markets with continuous transfers (e.g., \citealp{FJJT19,BEJK23,NV24,RY25}), within a matching-with-contracts framework (e.g., \citealp{R84,HM05}). These settings are also called nontransferable utility (NTU) markets and imperfectly transferable utility (ITU) markets, respectively. Under HIR, the regulator's preference is unrestricted in finite-contract markets. The existence result for markets with continuous transfers requires continuity and finite willingness to pay for all agents, including the regulator. HIR is weaker than substitutability in finite-contract markets and, under the stated monetary conditions, weaker than gross substitutability in markets with continuous transfers. It restricts preferences over residual bundles while allowing substantial flexibility in preferences over competing bundles. Under quasilinear utilities, HIR is equivalent to submodularity of the valuation.

Many agreements require the participation of an institution that authorizes, funds, or coordinates the activity. A public agency may join industrial production projects, a licensing authority may approve regulated projects, and a coordinating institution may participate in service agreements. We illustrate our HIR-based results with two applications. In manufacturing, companies undertake projects with fixed contractual obligations and face convex workload costs as production capacity becomes scarce. In centralized procurement of digital services, organizations acquire packages with overlapping functions through agreements involving a central purchasing institution and suppliers. Cash benefits are submodular, nonmonetary burdens are additive, and financial resources are evaluated through a concave utility function. In both applications, the regulator participates in every contract and ordinary agents satisfy HIR. Nevertheless, their preferences may violate substitutability in manufacturing and gross substitutability in procurement. The applications illustrate two mechanisms supporting HIR: increasing marginal workload costs and diminishing benefits from overlapping services.

We also establish a more general existence result for finite-contract markets. The regulator's preference satisfies \textbf{deletion alignment} (DEL) with respect to an ordinary agent if it supports every contract deletion that strictly benefits that agent. A setwise stable outcome exists if each ordinary agent either satisfies HIR or is covered by DEL. Thus agents may have arbitrary preferences provided that the regulator supports their profitable contract deletions. We apply this result to project staffing within a corporate group, where headquarters approves individual or collective worker assignments. Workers satisfy HIR through unit demand, while headquarters' evaluation of project returns ensures DEL with respect to the project units. 

Our framework also relates to the literature on matching under distributional constraints. This literature studies restrictions and policy goals concerning the distribution of workers or students across institutions, including regional caps (e.g., \citealp{KK15,KK17,KK18}), type-specific quotas (e.g., \citealp{A05,K12,EHHY14}), reserves (e.g., \citealp{HYY13,EY15,KHIY17}), minimum requirements (e.g., \citealp{BFIM10,FITUY15,FT17,Akin21}), proportionality targets (e.g., \citealp{NV19}), resource budgets (e.g., \citealp{NNT21,KK24,ABB24}), distributional objectives (e.g., \citealp{KK20,HKY25}), and general classes of constraints (e.g., \citealp{GKKTY17,KTY18}). Our framework offers an alternative perspective by giving the regulatory institution an explicit contractual role. For example, a regional cap can be represented by the regulator ranking every bundle that exceeds the cap below the empty bundle, while its ranking of acceptable bundles expresses further policy objectives. A setwise block must, however, make every participant, including the regulator, strictly better off. A deviation that respects a cap but leaves the regulator indifferent therefore does not block. Our existence result depends on this contractual role and blocking requirement.

Setwise stability has been formulated using several blocking conventions. \citet{R84} presents an early coalitional stability concept, and \citet{S99} explicitly formulates setwise stability in many-to-many matching. Their blocking tests differ, as explained by \citet{KW09}. Our definition follows the convention of \citet{EO06} and \citet{KW09}, and its multilateral formulation in \citet{BH21} and \citet{H24}: the deviators agree on a common final outcome that makes each of them strictly better off and gives each an individually rational bundle, while outsiders' residual bundles need not be individually rational. This last feature makes HIR central to our existence argument.

Our study also contributes to the broader literature on stability beyond standard substitutability assumptions. \citet{DKM01}, \citet{BK19}, and \citet{H23} study unimodularity conditions. \citet{SY06}, \citet{O08}, \citet{W10}, \citet{HK12}, \citet{HKNOW13,HKNOW19,HKNOW21}, and \citet{CEV21} study markets with same-group substitutability and cross-group complementarity, while \citet{HT26} studies same-group complementarity and cross-group substitutability. \citet{EY07}, \citet{P12}, \citet{RY20}, and \citet{PY23} examine peer effects and externalities among participants. The bilateral and unilateral substitutes conditions of \citet{HK10} relax substitutability and apply to cadet--branch matching (\citealp{S13,SS13}). \citet{HK15}, \citet{BH21}, and \citet{H24} study divisible participation, market structure, and weaker forms of complementarity in multilateral matching, respectively. \citet{AH18} and \citet{CKK19} examine stability in large markets. \citet{NV18,NV19,NV24} restore stability by slightly adjusting market parameters.

The remainder of the paper is organized as follows.
Section~\ref{sec:framework} presents the framework and selection principle.
Section~\ref{sec:ntu} studies finite-contract markets and an application
to manufacturing projects. Section~\ref{sec:tu} establishes existence
with continuous transfers and develops an application to centralized
procurement of digital services; Section~\ref{sec:quasi} establishes the equivalence between HIR and submodularity under quasilinear utilities. Section~\ref{sec:job} generalizes the existence result using deletion alignment and applies it to project staffing within a corporate group. Omitted proofs are provided in the appendix.

\section{Framework}\label{sec:framework}

\subsection{Preliminaries}

Let $I=\{r\}\cup J$ be a finite set of agents, where $r$ is the
regulator and $J$ is the set of ordinary agents.\footnote{The
regulator is not an ordinary agent: $r\notin J$.}
Let $X$ be a possibly infinite set of contracts. Each contract
$x\in X$ has a set of participants $\mathrm{N}(x)\subseteq I$
satisfying
\begin{equation}\label{eq:common}
 r\in\mathrm{N}(x),\qquad |\mathrm{N}(x)|\ge2.
\end{equation}
In particular, an ordinary agent may sign a bilateral contract
with the regulator. For $Y\subseteq X$, $i\in I$, and $K\subseteq I$,
write
\[
 \begin{aligned}
 Y_i&=\{x\in Y:i\in\mathrm{N}(x)\},&
 X_i&=\{x\in X:i\in\mathrm{N}(x)\},\\
 Y_K&=\bigcup_{i\in K}Y_i,&
 \mathrm{N}(Y)&=\bigcup_{x\in Y}\mathrm{N}(x).
 \end{aligned}
\]
Set $\mathrm{N}(\varnothing)=\varnothing$.

An \textbf{outcome} records the contracts that are signed.
In the finite-contract markets of Sections~\ref{sec:ntu}
and~\ref{sec:job}, the contract set $X$ is finite, and an outcome is any
set $Y\subseteq X$. In the market with continuous transfers in
Section~\ref{sec:tu}, an outcome is a finite
set $Y\subseteq X$ containing at most one contract for each
agreement.\footnote{An agreement is called a primitive contract
in \citet{RY20,RY25}.}
Agreements and their contract versions are defined in
Section~\ref{subsec:itusetting}.

A \textbf{bundle for agent $i$} is a finite set of contracts
$B\subseteq X_i$ that can occur together in an outcome.
Write $\mathcal B_i$ for the collection of these bundles.
In the finite-contract market, $\mathcal B_i=2^{X_i}$. With continuous
transfers, each bundle contains at most one contract for each
agreement. In either case, $\varnothing\in\mathcal B_i$, and
every subset of a bundle is a bundle.

Each agent $i\in I$ has a complete and transitive preference
relation $\succeq_i$ over $\mathcal B_i$, with strict part
$\succ_i$ and indifference relation $\sim_i$. There are no
externalities: each agent evaluates an outcome solely through
its own bundle. For a finite menu $Y\subseteq X$, define the
\textbf{choice correspondence} by
\begin{equation}\label{eq:choice}
 \mathcal C_i(Y)=
 \left\{
 B\in\mathcal B_i:
 \begin{array}{l}
 B\subseteq Y_i\text{ and }B\succeq_i B'\\
 \text{for every }B'\in\mathcal B_i\text{ with }B'\subseteq Y_i
 \end{array}
 \right\}.
\end{equation}
The choice set $\mathcal C_i(Y)$ is nonempty because the menu
is finite, the empty bundle is available, and preferences are
complete and transitive. In the finite-contract market of
Section~\ref{sec:ntu}, this definition reduces to the
conventional form
\[
 \mathcal C_i(Y)
 =\bigl\{
 B\subseteq Y_i:
 B\succeq_i B'\text{ for every }B'\subseteq Y_i
 \bigr\}.
\]
In the market with continuous transfers of Section~\ref{sec:tu}, a menu need not itself
be an outcome: it may offer several contract versions of the
same agreement, but each chosen bundle contains at most one
such version.

\subsection{Setwise stability}\label{Sec_stable}

An outcome $Y$ is \textbf{individually rational} for agent $i$
if $Y_i\in\mathcal C_i(Y)$. A bundle $B\in\mathcal B_i$ is
individually rational if $B\in\mathcal C_i(B)$. Thus individual
rationality rules out profitable unilateral deletions of
contracts. An outcome is individually rational if it is
individually rational for every agent. Define
\[
 \begin{aligned}
 \mathcal A_i
 &=\{B\in\mathcal B_i:B\in\mathcal C_i(B)\},\\
 \mathcal A
 &=\{Y:Y\text{ is an outcome and }Y_i\in\mathcal A_i
       \text{ for all }i\in I\}.
 \end{aligned}
\]

We use the setwise-blocking convention of
\citet[Definitions~4.1--4.2]{EO06} and \citet{KW09},
adapted to multilateral contracts as in
\citet[Definition~4]{BH21}. Agents entering new contracts
must agree on a common final outcome and receive individually
rational bundles. Individual rationality is not required of
the residual bundles of agents outside the blocking group.
\citet{KW09} distinguishes this convention from that of
\citet{S99}, which requires the blocking outcome to be
individually rational for all agents. This convention also
differs from the choice-based coalitional test in \citet{R84}.

\begin{definition}[Setwise stability]\label{Def_setwise}
An outcome $Y$ is \textbf{setwise blocked} by a nonempty
finite set $Z\subseteq X\setminus Y$ if there is a final
outcome $Y^*$ such that
\begin{equation}\label{eq:block}
 Z\subseteq Y^*\subseteq Y\cup Z,\qquad
 Y_i^*\succ_i Y_i,\quad Y_i^*\in\mathcal A_i
 \quad\text{for every }i\in\mathrm{N}(Z).
\end{equation}
An outcome is \textbf{setwise stable} if it is individually
rational and has no setwise block.
\end{definition}

The participants in a setwise block sign the contracts in $Z$
and may drop existing contracts. Each participant strictly
prefers its resulting bundle to its original bundle and finds
the resulting bundle individually rational. Outsiders'
residual bundles need not be individually rational. Since
the regulator signs every new contract, every setwise block
must make the regulator strictly better off.

For comparison, consider the finite-contract market with strict
preferences, following
\citet[Definitions~1 and~4, Examples~1 and~3]{BH21}.
An outcome $Y$ is \textbf{blocked} if some nonempty finite
set $Z\subseteq X\setminus Y$ satisfies
\begin{equation}\label{eq:bhblock}
 \forall i\in\mathrm{N}(Z)\quad
 \exists B_i\in\mathcal C_i(Y\cup Z):
 \quad Z_i\subseteq B_i.
\end{equation}
An outcome is \textbf{stable} if it is individually rational
and unblocked. This choice-based notion of stability goes
back to \citet[p.~50]{R84}. It tests agents' optimal choices
without requiring agreement on which existing contracts are
retained. Setwise blocking instead requires a common final
outcome in which each participant is strictly better off and
receives an individually rational bundle, without requiring
that bundle to be optimal from $Y\cup Z$.

Neither notion of stability implies the other, as illustrated
by the following two examples. Each example has two agents,
$r$ and $j$, both of whom participate in every
contract.\footnote{In the preference lists for finite-contract markets below, every unlisted bundle is strictly worse
than $\varnothing$.}

First, let $X=\{x,y\}$ and consider the preferences
\begin{equation}\label{exam_setwise}
 \begin{aligned}
 r:&\quad \{x,y\}\succ\{x\}\succ\varnothing,\\
 j:&\quad \{y\}\succ\{x\}\succ\varnothing.
 \end{aligned}
\end{equation}
The outcome $Y=\{x\}$ is blocked by $\{y\}$ under
\eqref{eq:bhblock}. However, the agents' optimal choices
disagree on whether $x$ should be retained: $r$ chooses
$\{x,y\}$, whereas $j$ chooses $\{y\}$. Neither $\{y\}$
nor $\{x,y\}$ makes both agents strictly better off.
Thus $Y$ is setwise stable.

Next, let $X=\{x,y,z\}$ and consider the preferences
\begin{equation}\label{eq:bh-threecontracts}
 \begin{aligned}
 r:&\quad \{x,y\}\succ\{x,z\}\succ
          \{y\}\succ\{z\}\succ\varnothing,\\
 j:&\quad \{x,z\}\succ\{y\}\succ\varnothing.
 \end{aligned}
\end{equation}
The outcome $\{y\}$ is stable but setwise blocked by
$\{x,z\}$. Both agents strictly prefer $\{x,z\}$ to
$\{y\}$, and $\{x,z\}$ is individually rational for
both agents. However, the regulator's optimal choice
from $\{x,y,z\}$ is $\{x,y\}$.

\subsection{Hereditary individual rationality}\label{Sec_HIR}

\begin{definition}[Hereditary individual rationality]\label{def:hir}
Agent $i$ satisfies \textbf{hereditary individual rationality}
(HIR) if, for every $S\in\mathcal B_i$ and every
$S'\subseteq S$,
\begin{equation}\label{eq:hir}
 S\in\mathcal C_i(S)
 \quad\Longrightarrow\quad
 S'\in\mathcal C_i(S').
\end{equation}
\end{definition}

Thus every subset of an individually rational bundle is
individually rational.\footnote{In the market with continuous transfers of
Section~\ref{sec:tu}, this condition applies to all bundles
in $\mathcal B_i$, not only to bundles occurring in outcomes
that are individually rational for every agent.}

\begin{definition}[Submodular set function]\label{def:submodularity}
Let $E$ be a set and $\mathcal S\subseteq2^E$ a family
closed under taking subsets. A function
$\mathrm{f}:\mathcal S\to\R\cup\{-\infty\}$ is
\textbf{submodular} if, for every $S,T\in\mathcal S$
such that $S\cup T\in\mathcal S$,
\begin{equation}\label{eq:submodularity}
 \mathrm{f}(S)+\mathrm{f}(T)
 \ge
 \mathrm{f}(S\cup T)+\mathrm{f}(S\cap T).
\end{equation}
A sum containing $-\infty$ is interpreted as $-\infty$.
\end{definition}

\begin{lemma}[A cardinal sufficient condition]\label{lem:submodular}
\normalfont
Suppose agent $i$'s preference admits a submodular
representation
\[
 \mathrm{u}_i:\mathcal B_i\to\R\cup\{-\infty\},
 \qquad
 \mathrm{u}_i(\varnothing)\in\R.
\]
Then the preference satisfies HIR. The converse is false,
even for strict preferences on a finite contract set.
\end{lemma}

Fix an individually rational bundle $A\in\mathcal A_i$.
Individual rationality implies
$\mathrm{u}_i(A)\ge\mathrm{u}_i(\varnothing)>-\infty$.
For every $B\subseteq A$, submodularity gives
\[
 \mathrm{u}_i(B)+\mathrm{u}_i(A\setminus B)
 \ge
 \mathrm{u}_i(A)+\mathrm{u}_i(\varnothing)>-\infty.
\]
Thus every subset of $A$ has finite utility.
For $D\subseteq B\subseteq A$, apply submodularity to
$B$ and $A\setminus(B\setminus D)$ to obtain
\[
 \mathrm{u}_i(B)-\mathrm{u}_i(D)
 \ge
 \mathrm{u}_i(A)
 -\mathrm{u}_i\bigl(A\setminus(B\setminus D)\bigr)
 \ge0.
\]
The last inequality follows from individual rationality
of $A$. In each application of submodularity, the union
of the two sets is a subset of $A$ and hence is a bundle.
Since $D$ is an arbitrary subset of $B$, the bundle $B$
is individually rational. This proves HIR.
The following example shows that the converse fails.

\begin{example}[HIR without a submodular representation]
\label{ex:hir-no-submodular-representation}
\normalfont
Let $X_i=\{x,y,z\}$, with $\mathcal B_i=2^{X_i}$,
and consider the strict preference
\begin{equation}\label{eq:hir-no-submodular-representation}
 \begin{aligned}
 &\{x\}\succ\{x,y,z\}\succ\{x,y\}\succ\{x,z\}\\
 &\hspace{1cm}\succ\{y,z\}\succ\{y\}\succ\{z\}\succ\varnothing.
 \end{aligned}
\end{equation}
The individually rational bundles are exactly
\[
 \mathcal A_i
 =\bigl\{
 \varnothing,\{x\},\{y\},\{z\},\{y,z\}
 \bigr\}.
\]
Since $\mathcal A_i$ is closed under taking subsets,
the preference satisfies HIR.

Nevertheless, no utility representation of this preference
is submodular. Any representation $\mathrm{u}_i$ must satisfy
\[
 \mathrm{u}_i(\{x\})>\mathrm{u}_i(\{x,y\}),
 \qquad
 \mathrm{u}_i(\{x,y,z\})>\mathrm{u}_i(\{x,z\}).
\]
Adding these inequalities gives
\[
 \mathrm{u}_i(\{x\})+\mathrm{u}_i(\{x,y,z\})
 >
 \mathrm{u}_i(\{x,y\})+\mathrm{u}_i(\{x,z\}),
\]
contradicting submodularity applied to $\{x,y\}$ and
$\{x,z\}$. HIR restricts preferences over subsets of
individually rational bundles, whereas a submodular
representation also restricts comparisons involving
bundles that are not individually rational.
\end{example}

\begin{proposition}[The regulator's selection principle]
\label{lem:selection}
\normalfont
Suppose every ordinary agent satisfies HIR. If an outcome
$Y^r\in\mathcal A$ is weakly preferred by the regulator
to every other outcome in $\mathcal A$, then $Y^r$ is
setwise stable.
\end{proposition}

The regulator's participation is central to the argument:
every setwise block must make it strictly better off.
Suppose $Y^r$ is setwise blocked by $Z$ with final outcome
$Y^*$. The participants' final bundles are individually
rational by the definition of a setwise block.
Since the regulator participates in every block, every
outsider is an ordinary agent. An outsider can only lose
contracts, so HIR ensures that its residual bundle is
individually rational. Thus $Y^*\in\mathcal A$, but the
regulator strictly prefers $Y^*$ to $Y^r$, contradicting
the choice of $Y^r$.

\begin{example}[Nonexistence when an ordinary agent fails HIR]
\label{ex:hir-nonexistence}
Let $I=\{r,i,j\}$ and $X=\{x,y,z\}$, with
\[
 \mathrm{N}(x)=\mathrm{N}(y)=\{r,i\},
 \qquad
 \mathrm{N}(z)=\{r,j\}.
\]
Preferences are strict, and the acceptable bundles are
ranked as follows:
\begin{equation}\label{eq:hir-nonexistence}
 \begin{aligned}
 r:&\quad \{x,z\}\succ\{x,y\}\succ\{z\}\succ\varnothing,\\
 i:&\quad \{x,y\}\succ\varnothing,\\
 j:&\quad \{z\}\succ\varnothing.
 \end{aligned}
\end{equation}
Among the ordinary agents, only $i$ fails HIR:
its bundle $\{x,y\}$ is individually rational, but
neither singleton is individually rational.

The individually rational outcomes are exactly
\[
 \mathcal A=\{\varnothing,\{z\},\{x,y\}\}.
\]
None is setwise stable.

The empty outcome is setwise blocked by
$Z=Y^*=\{z\}$, since both $r$ and $j$ strictly prefer
$\{z\}$ to $\varnothing$ and find it individually rational.

The outcome $\{z\}$ is setwise blocked by
$Z=Y^*=\{x,y\}$. Both $r$ and $i$ are strictly better
off and receive individually rational bundles.
Although $\{x,y\}$ is not the regulator's optimal
choice from $\{x,y,z\}$, this does not prevent a
setwise block.

The outcome $Y=\{x,y\}$ is setwise blocked by
$Z=\{z\}$ with $Y^*=\{x,z\}$. The participants are
$r$ and $j$. The regulator strictly prefers
$\{x,z\}$ to $\{x,y\}$ and finds $\{x,z\}$
individually rational, while $j$ strictly prefers
$\{z\}$ to its previous empty bundle. Outsider $i$
loses $y$ and retains the unacceptable bundle $\{x\}$.

Thus $\{x,y\}$ is the regulator's most preferred
outcome in $\mathcal A$, but the final outcome
$\{x,z\}$ of its setwise block lies outside
$\mathcal A$. The failure of HIR prevents the
selection argument from applying.

To illustrate the role of this failure, change only
$i$'s preference to
$\{x,y\}\succ\{x\}\succ\{y\}\succ\varnothing$.
Then $i$ satisfies HIR, and $\{x,z\}$ is individually
rational for every agent and is the regulator's most
preferred outcome. It is therefore setwise stable
by Proposition~\ref{lem:selection}.
\end{example}

\section{Finite-contract markets}\label{sec:ntu}

Throughout this section, the contract set $X$ is finite,
and an outcome is any set $Y\subseteq X$.
Contracts specify fixed obligations, which may include
payments or receipts.

\begin{corollary}[Existence in finite-contract markets]
\label{thm:hub}
\normalfont
If every ordinary agent satisfies HIR, a setwise stable
outcome exists. Every outcome most preferred by the
regulator among those in $\mathcal A$ is setwise stable.
\end{corollary}

The family $\mathcal A$ is finite and contains the empty
outcome. The regulator therefore has a most preferred
outcome in $\mathcal A$. Proposition~\ref{lem:selection}
then yields both conclusions. The regulator's preference
may be any complete and transitive relation, with
indifferences allowed.

\subsection{Substitutability}

We compare substitutability with hereditary individual rationality in finite-contract markets, allowing for indifferences.

\begin{definition}[Substitutability with indifferences]
\label{def:substitutes}
Agent $i$ has a substitutable preference if, for every
$S'\subseteq S\subseteq X_i$, both of the following
conditions hold:
\begin{align}
 &\forall A\in\mathcal C_i(S)\quad
  \exists B\in\mathcal C_i(S'):
  \quad A\cap S'\subseteq B;
  \label{eq:s1}\tag{S1}\\
 &\forall B\in\mathcal C_i(S')\quad
  \exists A\in\mathcal C_i(S):
  \quad A\cap S'\subseteq B.
  \label{eq:s2}\tag{S2}
\end{align}
Condition~\eqref{eq:s1} is \textbf{admission monotonicity},
and condition~\eqref{eq:s2} is \textbf{rejection monotonicity}.
\end{definition}

For each optimal bundle from the larger menu, admission
monotonicity requires an optimal bundle from the smaller
menu that retains every contract in the original bundle
that remains available. For each optimal bundle from the
smaller menu, rejection monotonicity requires an optimal
bundle from the larger menu that continues to reject every
contract rejected from the smaller menu.

Under strict preferences, the two conditions are equivalent
and reduce to the conventional definition of substitutability
in \citet{RS90}. The formulation allowing indifferences is
presented in \citet{S99}. With indifferences, neither
condition implies the other.

\begin{proposition}\label{prop:sub-hir}
\normalfont
In finite-contract markets:
\begin{description}
 \item[(i)] Admission monotonicity implies HIR.
 The converse fails even under strict preferences.

 \item[(ii)] Rejection monotonicity and HIR are logically
 independent when indifferences are allowed.\footnote{
 Under strict preferences, rejection monotonicity is
 equivalent to admission monotonicity, so part~(i) also
 describes its relationship with HIR.}
\end{description}
\end{proposition}

To show that admission monotonicity implies HIR, suppose
$A\in\mathcal C_i(A)$ and $B\subseteq A$. By admission
monotonicity, there exists $D\in\mathcal C_i(B)$ such that
\[
 A\cap B=B\subseteq D\subseteq B.
\]
Hence $D=B$, so $B\in\mathcal C_i(B)$. This proves HIR.

\begin{example}[HIR with strict preferences, but without admission monotonicity]
\label{ex:hir-nosubstitute}
\normalfont
Let $X_i=\{x,y,z\}$ and consider the strict preference
\begin{equation}\label{eq:ordinal-example}
 \{x,y\}\succ\{z\}\succ\{y\}\succ\{x\}\succ\varnothing.
\end{equation}
The individually rational bundles are exactly
\[
 \mathcal A_i
 =\bigl\{
 \varnothing,\{x\},\{y\},\{z\},\{x,y\}
 \bigr\}.
\]
This family is closed under taking subsets, so the
preference satisfies HIR. Nevertheless,
\[
 \mathcal C_i(\{x,y,z\})=\bigl\{\{x,y\}\bigr\},
 \qquad
 \mathcal C_i(\{x,z\})=\bigl\{\{z\}\bigr\}.
\]
When $y$ is removed from the menu, the agent no longer
chooses $x$, even though $x$ remains available. This
violates admission monotonicity.

HIR concerns subsets of individually rational bundles,
whereas admission monotonicity concerns how optimal
choices respond to reductions in the available menu.
\end{example}

\begin{example}[Rejection monotonicity without HIR]
\label{ex:rm-nohir}
Let $X_i=\{x,y\}$ and consider the preference
\begin{equation}\label{eq:s2-not-hir}
\{x,y\}\sim_i\varnothing\succ_i \text{ any other bundle }.
\end{equation}
As stipulated in Section~\ref{Sec_stable}, every unlisted
bundle is strictly worse than $\varnothing$.
The empty bundle is therefore optimal from every menu:
\[
 \varnothing\in\mathcal C_i(S)
 \qquad\text{for all }S\subseteq X_i.
\]
For every pair $S'\subseteq S$ and every
$B\in\mathcal C_i(S')$, choose
$A=\varnothing\in\mathcal C_i(S)$. Then
\[
 A\cap S'=\varnothing\subseteq B,
\]
so rejection monotonicity holds. However, $\{x,y\}$ is
individually rational, whereas its subset $\{x\}$ is not.
Thus HIR fails.

Conversely, Example~\ref{ex:hir-nosubstitute} shows that
HIR does not imply rejection monotonicity, because
admission monotonicity and rejection monotonicity are
equivalent under strict preferences.
\end{example}

\subsection{Application: manufacturing projects}
\label{subsec:ntuapplication}

Consider manufacturing companies choosing which joint
production projects to undertake under a regulatory
program. A project may involve several companies
contributing components, specialized equipment, or
different stages of production. Each contract specifies
the project, its participating companies, their fixed
production commitments, and any fixed subsidies or payment
obligations. A regulatory institution $r$, such as a public
agency coordinating an industrial production program,
participates in every contract and oversees the approval
and coordination of projects. Its preference may reflect
product quality, environmental standards, and the
allocation of support across projects.

Projects may generate different benefits for a company
and require different amounts of production capacity.
Benefits may reflect production revenue net of
project-specific costs and fixed contractual payments.
Workload may be measured in labor hours or machine time.
We consider production with
\textbf{nondecreasing marginal workload costs}.
When staffing and equipment are fixed, initial projects
can use idle workers and spare production capacity at
relatively low cost. As workload rises, additional projects
may require overtime, extra shifts, more intensive
supervision, or the rescheduling of production.
Congestion, fatigue, and accelerated equipment wear
may further increase marginal workload costs.
A convex workload cost function captures these aggregate
costs beyond the project-specific costs already included
in each project's benefit.

For each manufacturing company $i\in J$ and contract
$x\in X_i$, let $v_{ix}>0$ denote the company's benefit
from $x$ and $r_{ix}\ge0$ the production workload required
by $x$. Let
$\mathrm{g}_i:\mathbb R_+\to\mathbb R$
be nondecreasing and convex, with $\mathrm{g}_i(0)=0$.
Define
\begin{equation}\label{eq:workload-utility}
 \mathrm{R}_i(S)=\sum_{x\in S}r_{ix},
 \qquad
 \mathrm{u}_i(S)
 =\sum_{x\in S}v_{ix}
  -\mathrm{g}_i\bigl(\mathrm{R}_i(S)\bigr),
 \qquad S\subseteq X_i.
\end{equation}
The company's preference is represented by $\mathrm{u}_i$.
For any menu $M\subseteq X$, its choice correspondence is
\[
 \mathcal C_i(M)
 =\operatorname*{arg\,max}_{S\subseteq M_i}
   \mathrm{u}_i(S).
\]

\begin{proposition}[Convex workload costs imply HIR]
\label{prop:workload-hir}
\normalfont
Every preference represented by
\eqref{eq:workload-utility} satisfies HIR.
This class includes preferences that violate substitutability.
\end{proposition}

Convexity of $\mathrm{g}_i$ and nonnegative workloads
imply that the incremental workload cost of a given
project weakly increases as additional projects are
undertaken. Its marginal utility therefore weakly
decreases, so $\mathrm{u}_i$ is submodular.
Since $\mathrm{u}_i(\varnothing)=0$,
Lemma~\ref{lem:submodular} implies HIR.
The following example establishes the failure of
substitutability.

\begin{example}[HIR with indifferences, but without substitutability]
\label{ex:workload-nonsubstitutable}
Let $X_i=\{x,y,z\}$ and set
\[
 (v_{ix},v_{iy},v_{iz})=(2,2,3),
 \qquad
 (r_{ix},r_{iy},r_{iz})=(1,2,3),
 \qquad
 \mathrm{g}_i(R)=10\max\{0,R-3\}.
\]
The company has three units of regular production
capacity and incurs a cost of ten for each unit of
workload beyond that capacity. The utilities of the
acceptable bundles are:
\begin{center}
\begin{tabular}{@{}lrrrrr@{}}
\toprule
$S$
 & $\varnothing$
 & $\{x\}$
 & $\{y\}$
 & $\{z\}$
 & $\{x,y\}$\\
\midrule
$\mathrm{u}_i(S)$
 & $0$ & $2$ & $2$ & $3$ & $4$\\
\bottomrule
\end{tabular}
\end{center}
Thus the preference list is
\[
 \{x,y\}\succ_i\{z\}\succ_i
 \{x\}\sim_i\{y\}\succ_i\varnothing.
\]
The family of individually rational bundles is
\[
 \mathcal A_i
 =\bigl\{
 \varnothing,\{x\},\{y\},\{z\},\{x,y\}
 \bigr\},
\]
which is closed under taking subsets.
Nevertheless, for $S=\{x,y,z\}$ and $S'=\{x,z\}$,
\[
 \mathcal C_i(S)=\bigl\{\{x,y\}\bigr\},
 \qquad
 \mathcal C_i(S')=\bigl\{\{z\}\bigr\}.
\]

Condition~\eqref{eq:s1} fails because the unique optimal
bundle from $S$ is $A=\{x,y\}$, but no optimal bundle
from $S'$ contains $A\cap S'=\{x\}$.
Condition~\eqref{eq:s2} also fails: for the unique
optimal bundle $B=\{z\}$ from $S'$, the only optimal
bundle from $S$ is $A=\{x,y\}$, and
$A\cap S'=\{x\}\not\subseteq B$.
The preference therefore satisfies HIR while violating
both substitutability conditions.\footnote{
The same structure also yields an example with strict
preferences. Keeping workloads and costs fixed but
setting $(v_{ix},v_{iy},v_{iz})=(2,3,4)$ makes the
preference strict and gives the list in
\eqref{eq:ordinal-example}. The same two menus
violate substitutability.}
\end{example}

\begin{remark}[Several resources]\label{rem:workload-resources}
The same argument applies when projects use several
resources:
\[
 \mathrm{u}_i(S)
 =\sum_{x\in S}v_{ix}
  -\sum_{k=1}^{d_i}
   \mathrm{g}_{ik}\!
   \left(\sum_{x\in S}r_{ikx}\right),
\]
where $d_i$ is the number of resources, $r_{ikx}\ge0$,
and each $\mathrm{g}_{ik}$ is nondecreasing and convex,
with $\mathrm{g}_{ik}(0)=0$.
Resource requirements may be measured in labor hours,
machine time, or warehouse space.
\end{remark}

\section{Markets with continuous transfers}\label{sec:tu}

We now allow continuous transfers, following the notation
of \citet{RY25}. Utilities need not be quasilinear.
Section~\ref{sec:quasi} specializes this model to quasilinear utilities.

\subsection{Setting}\label{subsec:itusetting}

Let $\Omega$ be a finite set of agreements.
Each agreement $w\in\Omega$ has a set of signatories
$\mathrm{N}(w)$ satisfying \eqref{eq:common}.
A contract $x=(w,\mathbf t^w)$ specifies an agreement
and a balanced transfer vector:
\begin{equation}\label{eq:balance}
 \mathbf t^w=(t_i^w)_{i\in\mathrm{N}(w)}
 \in\R^{\mathrm{N}(w)},
 \qquad
 \sum_{i\in\mathrm{N}(w)}t_i^w=0.
\end{equation}
A positive $t_i^w$ denotes a payment by agent $i$,
and a negative value denotes a receipt.
The contract set is
\[
 X=
 \left\{
 (w,\mathbf t^w):
 w\in\Omega,\ 
 \mathbf t^w\in\R^{\mathrm{N}(w)},\
 \sum_{i\in\mathrm{N}(w)}t_i^w=0
 \right\}.
\]

Write $\mathrm{\uptau}(x)=w$ for the agreement specified
by $x=(w,\mathbf t^w)$, and let
$\mathrm{\uptau}(Y)=\{\mathrm{\uptau}(x):x\in Y\}$.
An \textbf{outcome} is a finite set $Y\subseteq X$
such that
\[
 \mathrm{\uptau}(x)\ne\mathrm{\uptau}(x')
 \quad\text{for all distinct }x,x'\in Y.
\]
For $\Psi\subseteq\Omega$ and $i\in I$, let
\[
 \Psi_i=\{w\in\Psi:i\in\mathrm{N}(w)\},
 \qquad
 \Omega_i=\{w\in\Omega:i\in\mathrm{N}(w)\}.
\]
A subset of $\Omega_i$ is a
\textbf{portfolio of agreements} for agent $i$.
A \textbf{bundle of contracts} $A\in\mathcal B_i$
is a finite subset of $X_i$ containing at most one
contract for each agreement.

For each agent $i$, let
\[
 \mathrm{V}_i:
 2^{\Omega_i}\times\R\to\R\cup\{-\infty\},
 \qquad
 \mathrm{V}_i(\varnothing,0)\in\R.
\]
The agent's preference over bundles is represented by
\begin{equation}\label{eq:utility}
 \mathrm{u}_i(Y_i)
 =
 \mathrm{V}_i\left(
 \mathrm{\uptau}(Y_i),
 -\sum_{(w,\mathbf t^w)\in Y_i}t_i^w
 \right).
\end{equation}
The monetary argument is a scalar equal to total net
receipts. Any fixed initial wealth is incorporated in
$\mathrm{V}_i$, and the monetary argument is zero in
the absence of contracts. Money is fungible across
an agent's agreements, but utility need not be
additively separable or linear in money.

The value $-\infty$ can represent technological
constraints. For example, a technological domain
$\mathcal D_i\subseteq2^{\Omega_i}$ with
$\varnothing\in\mathcal D_i$ can be represented by
setting $\mathrm{V}_i(\Psi,z)=-\infty$ for every $z$
when $\Psi\notin\mathcal D_i$, while taking
$\mathrm{V}_i(\Psi,\cdot)$ to be real-valued when
$\Psi\in\mathcal D_i$.

Because the regulator participates in every agreement,
it can rule out specified portfolios by assigning
$\mathrm{V}_r(\Psi,z)=-\infty$ for every $z$ on
those portfolios.

\subsection{Conditions on monetary preferences}
\label{subsec:monetaryconditions}

\begin{definition}[Monetary preference conditions]
\label{def:monetaryconditions}
\normalfont
Agent $i$ satisfies the
\textbf{monetary preference conditions} if
$\mathrm{V}_i(\varnothing,0)\in\R$ and, for every
$\Psi\subseteq\Omega_i$, the following conditions hold:
\begin{enumerate}
 \item \textbf{Continuity:}
 $\mathrm{V}_i(\Psi,\cdot)$ is continuous when
 $\R\cup\{-\infty\}$ is endowed with its order topology.
 Equivalently,
 $z\mapsto\exp(\mathrm{V}_i(\Psi,z))$
 is a continuous real-valued function, with the
 convention $\exp(-\infty)=0$.

 \item \textbf{Finite willingness to pay:}
 There exists a finite number $\mathrm{b}_i(\Psi)$
 such that, for every $P\in\R$,
 \begin{equation}\label{eq:fwtp}
  \mathrm{V}_i(\Psi,-P)
  \ge\mathrm{V}_i(\varnothing,0)
  \quad\Longrightarrow\quad
  P\le\mathrm{b}_i(\Psi).
 \end{equation}
\end{enumerate}
\end{definition}

Conditions similar to \eqref{eq:fwtp} have been used in \citet{FJJT19}, \citet{BEJK23}, and \citet{NV24}.

In the specification based on a technological domain
$\mathcal D_i$, continuity reduces to ordinary continuity
of $\mathrm{V}_i(\Psi,\cdot)$ for $\Psi\in\mathcal D_i$.
If utility can become $-\infty$ as the monetary argument
varies, continuity is required for the entire extended-real
function, not merely on its finite-valued domain.
For example, assigning $-\infty$ below a monetary
threshold and a finite utility at the threshold
violates this continuity condition.

The bounds $\mathrm{b}_i(\Psi)$ in \eqref{eq:fwtp}
may be enlarged to be nonnegative.
A sufficient condition for \eqref{eq:fwtp} is
$\mathrm{V}_i(\Psi,z)\to-\infty$ as $z\to-\infty$.
More generally, it suffices that this limit exists
and lies strictly below $\mathrm{V}_i(\varnothing,0)$.
If $\mathrm{V}_i(\Psi,\cdot)\equiv-\infty$,
the antecedent of \eqref{eq:fwtp} is always false,
so any finite bound may be used.

These conditions do not impose monetary monotonicity:
$\mathrm{V}_i(\Psi,\cdot)$ need not be increasing in money.

\begin{lemma}[Compactness of individually rational transfers]
\label{lem:compact}
\normalfont
Suppose every agent, including the regulator, satisfies
the monetary preference conditions, and every ordinary
agent satisfies HIR. Fix a set of executed agreements
$\Psi\subseteq\Omega$. The set of balanced transfer
vectors implementing an individually rational outcome
with agreement set $\Psi$ is a possibly empty compact
subset of the Euclidean space
$\prod_{w\in\Psi}\R^{\mathrm{N}(w)}$.\footnote{
Only transfers for agreements in $\Psi$ are coordinates
of this space.}
\end{lemma}

\begin{proposition}[Existence without quasilinearity]
\label{thm:main}
\normalfont
In the model of Section~\ref{subsec:itusetting},
suppose every agent, including the regulator, satisfies
the monetary preference conditions, and every ordinary
agent satisfies HIR. Then a setwise stable outcome exists.
More specifically, the regulator's utility attains a
maximum over all individually rational outcomes, and
every such maximizer is setwise stable. All agents have
finite utility at every individually rational outcome.
\end{proposition}

The empty outcome belongs to $\mathcal A$, and there
are finitely many possible sets of executed agreements.
For each such set with a nonempty individually rational
transfer set, Lemma~\ref{lem:compact} and continuity
ensure that $\exp(\mathrm{u}_r)$ attains a maximum.
The largest of these finitely many maxima is a global
maximum over $\mathcal A$. Since the exponential
transformation is strictly increasing on
$\R\cup\{-\infty\}$, this also maximizes
$\mathrm{u}_r$.

At any individually rational outcome $Y$, each agent
can delete all of its contracts, so
\[
 \mathrm{u}_i(Y_i)
 \ge\mathrm{V}_i(\varnothing,0)>-\infty.
\]
Every agent's utility is therefore finite.
Proposition~\ref{lem:selection} establishes setwise
stability of every outcome that maximizes the
regulator's utility over $\mathcal A$.

No HIR, substitutability, or submodularity condition
is imposed on the regulator. In particular, the
portfolios that can yield finite utility for the
regulator need not form a family closed under
taking subsets.

\subsection{Gross substitutability}\label{subsec:gs}

Fix an agent $i$. A vector
$\mathbf p_i=(p_i^w)_{w\in\Omega_i}\in\R^{\Omega_i}$
specifies the agent's payments, and
\[
 \mathrm{p}_i(\Psi)=\sum_{w\in\Psi}p_i^w
\]
is its net payment for portfolio $\Psi$.
Write $\mathbf{0}=(0)_{w\in\Omega_i}$ for the zero
payment vector. Scalar zero is denoted by $0$,
as in $\mathrm{V}_i(\varnothing,0)$.
Define
\begin{equation}\label{eq:gsdemand}
 \mathrm{D}_i(\mathbf p_i)
 =
 \argmax_{\Psi\subseteq\Omega_i}
 \mathrm{V}_i(\Psi,-\mathrm{p}_i(\Psi)).
\end{equation}
The demand set $\mathrm{D}_i(\mathbf p_i)$ is nonempty,
and every demanded portfolio has finite utility.
The payment vectors are used to formulate individual
preference conditions, without imposing market clearing.

\begin{definition}[Gross substitutability]\label{def:gs}
Agreements are \textbf{gross substitutes} (GS) for
agent $i$ if, for every
$\mathbf p_i,\mathbf q_i\in\R^{\Omega_i}$ with
$\mathbf p_i\le\mathbf q_i$ coordinatewise and every
$\Psi\in\mathrm{D}_i(\mathbf p_i)$,
there exists $\Psi'\in\mathrm{D}_i(\mathbf q_i)$
such that
\begin{equation}\label{eq:gs}
 \Psi\cap
 \{w\in\Omega_i:p_i^w=q_i^w\}
 \subseteq\Psi'.
\end{equation}
\end{definition}

For a fixed payment vector $\mathbf p_i$ and a menu
$\Psi\subseteq\Omega_i$, define
\begin{equation}\label{eq:gsmenu}
 \mathcal C_i(\Psi;\mathbf p_i)
 =
 \argmax_{\Phi\subseteq\Psi}
 \mathrm{V}_i(\Phi,-\mathrm{p}_i(\Phi)).
\end{equation}
This correspondence selects portfolios of agreements,
whereas $\mathcal C_i(Y)$ in \eqref{eq:choice} selects
bundles of contracts. Here each available agreement
has one specified payment.

Agent $i$ satisfies \textbf{monetary admission monotonicity}
if, for every $\mathbf p_i\in\R^{\Omega_i}$ and every
pair of menus $\Psi'\subseteq\Psi\subseteq\Omega_i$,
\begin{equation}\label{eq:gssone}
 \begin{gathered}
 \Psi'\subseteq\Psi,\quad
 \Phi\in\mathcal C_i(\Psi;\mathbf p_i)
 \\
 \Longrightarrow\quad
 \exists\Phi'\in\mathcal C_i(\Psi';\mathbf p_i):
 \quad \Phi\cap\Psi'\subseteq\Phi'.
 \end{gathered}
\end{equation}

\begin{proposition}[GS, admission monotonicity, and HIR]
\label{prop:gshir}
\normalfont
For agent $i$ in the model of
Section~\ref{subsec:itusetting}:
\begin{enumerate}[label=(\roman*)]
 \item If the outside option is finite and finite
 willingness to pay \eqref{eq:fwtp} holds for every
 portfolio, then GS implies monetary admission
 monotonicity \eqref{eq:gssone}. This implication
 does not require continuity.

 \item Monetary admission monotonicity implies HIR.
 This implication does not require continuity or
 finite willingness to pay.
\end{enumerate}
Consequently, under the assumptions in part~(i),
\[
 \mathrm{GS}
 \Longrightarrow
 \text{monetary admission monotonicity}
 \Longrightarrow
 \HIR.
\]
\end{proposition}

The following application provides a class of preferences
that satisfy HIR but need not satisfy monetary admission
monotonicity.

\subsection{Application: centralized procurement of digital services}
\label{subsec:ituapplication}

Consider public organizations procuring packages of
digital services through a central purchasing institution.
Packages may provide document management, accounting,
reporting, or data-analysis functions, and different
packages may cover overlapping functions.
Central purchasing bodies acquire services or award
contracts for other public organizations
(\citealp{OECD25}), while duplicate or unnecessary
software licenses are a documented concern in public
procurement (\citealp{GAO25}).
These features motivate the following model.

Each agreement specifies a service package, its suppliers
and user organizations, and their service obligations.
We consider a procurement arrangement in which the central
purchasing institution $r$ is a signatory to every agreement,
together with the relevant suppliers and users.
The institution approves the agreements and their transfers.
Its preference may reflect cybersecurity, interoperability,
expenditure, and the distribution of services across
organizations. Each contract additionally specifies
a transfer vector among the signatories, satisfying
\eqref{eq:balance}.

For a user organization $i$, let $K_i$ be a finite set
of functions and $K_{iw}\subseteq K_i$ the set of functions
delivered to it by agreement $w\in\Omega_i$.
Function $k$ generates cash savings $a_{ik}\ge0$,
such as avoided expenditure on a separately purchased
service. Suppose that obtaining the same function
through another package generates no additional savings.
The cash benefit of a portfolio is then
\begin{equation}\label{eq:procurement-coverage}
 \mathrm{v}_i(\Psi)
 =
 \sum_{k\in\bigcup_{w\in\Psi}K_{iw}}a_{ik},
 \qquad \Psi\subseteq\Omega_i.
\end{equation}
This function is submodular: the incremental savings
from a package weakly decrease as more of its functions
are already covered.

Each agreement also entails a nonmonetary burden,
such as the attention required to administer the
service and oversee its use. Let $c_{iw}\ge0$ denote
this burden, measured in utility units.
We assume that these burdens are additive across
agreements.

More generally, for each ordinary agent $i\in J$, let
$\mathrm{v}_i:2^{\Omega_i}\to\R$ be submodular, with
$\mathrm{v}_i(\varnothing)=0$, and let $c_{iw}\ge0$
for every $w\in\Omega_i$.
The valuation need not be nondecreasing under set
inclusion. For example, a supplier that incurs a cash
cost $k_{iw}\ge0$ of providing the service under
agreement $w$ has the modular valuation
\[
 \mathrm{v}_i(\Psi)=-\sum_{w\in\Psi}k_{iw}.
\]
Subtracting additive monetary implementation costs
from \eqref{eq:procurement-coverage} also preserves
submodularity.

Let $\mathrm{h}_i:\R\to\R$ be continuous, strictly
increasing, and concave, with $\mathrm{h}_i(0)=0$.
Define
\begin{equation}\label{eq:procurement-income}
 \mathrm{V}_i(\Psi,z)
 =
 \mathrm{h}_i\bigl(z+\mathrm{v}_i(\Psi)\bigr)
 -\sum_{w\in\Psi}c_{iw},
 \qquad
 \Psi\subseteq\Omega_i,\quad z\in\R.
\end{equation}
Here $z$ denotes total net receipts from contractual
transfers, so $z+\mathrm{v}_i(\Psi)$ is the net
financial contribution of the agreements.
Cash expenditures are included in $\mathrm{v}_i$
or the transfers, while the terms $c_{iw}$ measure
nonmonetary burdens.

Concavity of $\mathrm{h}_i$ represents
\textbf{diminishing marginal utility of financial resources}.
An organization can use additional funds for other
activities, allocating the first available units to
its most pressing needs. Once these needs are met,
further funds may have lower marginal value.
For an owner-managed supplier, the same condition
can describe diminishing marginal utility of business
income. Any fixed financial resources are incorporated
in $\mathrm{h}_i$.

When $\mathrm{h}_i$ is strictly concave, differences
in nonmonetary burdens can generate income effects.
In particular, at a financial resource level $m$,
the cash compensation $q>0$ for a fixed additional
burden $d>0$ satisfies
\[
 \mathrm{h}_i(m+q)-\mathrm{h}_i(m)=d.
\]
Whenever such compensation exists, the required $q$
increases with $m$. Because the nonmonetary burden
enters utility outside $\mathrm{h}_i$, this specification
need not represent a quasilinear preference.

\begin{proposition}[Centralized procurement]
\label{prop:procurement-income-hir}
\normalfont
Under the assumptions stated above, every preference
represented by \eqref{eq:procurement-income} satisfies
the monetary preference conditions and HIR.
This class includes preferences that violate monetary
admission monotonicity \eqref{eq:gssone}, and hence GS.
If every ordinary agent has such preferences and the
regulator satisfies the monetary preference conditions,
a setwise stable outcome exists.
\end{proposition}

\begin{example}[Overlapping packages without monetary admission monotonicity]
\label{ex:procurement-income-nongs}
Consider a user organization with three agreements
$\Omega_i=\{w_1,w_2,w_3\}$ covering four functions.
Let
\[
 K_{iw_1}=\{1,2\},
 \qquad
 K_{iw_2}=\{3,4\},
 \qquad
 K_{iw_3}=\{2,3\},
\]
and set
\[
 (a_{i1},a_{i2},a_{i3},a_{i4})
 =
 \left(\tfrac12,\tfrac32,\tfrac32,\tfrac12\right),
 \qquad
 c_{iw}=\tfrac34
 \quad(w\in\Omega_i).
\]
Use the coverage valuation
\eqref{eq:procurement-coverage} and let
\begin{equation}\label{eq:procurement-example}
 \mathrm{h}_i(m)=m-e^{-m}+1.
\end{equation}
We have
$\mathrm{h}_i(0)=0$,
$\mathrm{h}_i'(m)=1+e^{-m}>0$, and
$\mathrm{h}_i''(m)=-e^{-m}<0$.
The preference therefore satisfies HIR by
Proposition~\ref{prop:procurement-income-hir}.

At zero payments, the table below reports portfolio
valuations and utilities, with utilities rounded
to three decimal places.
\begin{center}
\begin{tabular}{@{}lrr@{}}
\toprule
$\Psi$
 & $\mathrm{v}_i(\Psi)$
 & $\mathrm{V}_i(\Psi,0)$\\
\midrule
$\varnothing$
 & $0$ & $0$\\
$\{w_1\}$, $\{w_2\}$
 & $2$ & $2.115$\\
$\{w_3\}$
 & $3$ & $3.200$\\
$\{w_1,w_2\}$
 & $4$ & $3.482$\\
$\{w_1,w_3\}$, $\{w_2,w_3\}$
 & $3.5$ & $2.970$\\
$\{w_1,w_2,w_3\}$
 & $4$ & $2.732$\\
\bottomrule
\end{tabular}
\end{center}
The strict comparisons follow directly from
\[
 \mathrm{V}_i(\Psi,0)
 =
 \mathrm{v}_i(\Psi)
 -e^{-\mathrm{v}_i(\Psi)}
 +1-\tfrac34|\Psi|.
\]
In particular,
\[
 \mathcal C_i(\Omega_i;\mathbf{0})
 =
 \bigl\{\{w_1,w_2\}\bigr\},
 \qquad
 \mathcal C_i(\{w_1,w_3\};\mathbf{0})
 =
 \bigl\{\{w_3\}\bigr\}.
\]
When $w_2$ is removed from the menu, the organization
no longer chooses $w_1$, even though $w_1$ remains
available. This violates monetary admission monotonicity.
Since the monetary preference conditions hold,
Proposition~\ref{prop:gshir}(i) also implies that GS fails.
\end{example}

Packages $w_1$ and $w_2$ jointly cover all four functions,
whereas $w_3$ covers the two most valuable functions.
When $w_2$ is unavailable, adding $w_1$ to the portfolio
$\{w_3\}$ delivers only the low-value function~1 and
incurs an additional nonmonetary burden.

\subsection{Quasilinear utilities}\label{sec:quasi}

Suppose every ordinary agent has quasilinear utility
\begin{equation}\label{eq:quasilinear}
 \mathrm{V}_i(\Psi,z)=\mathrm{v}_i(\Psi)+z,
 \qquad
 \mathrm{v}_i:2^{\Omega_i}\to\R\cup\{-\infty\},
 \quad
 \mathrm{v}_i(\varnothing)\in\R,
\end{equation}
with the convention $-\infty+z=-\infty$.
The regulator's utility need not be quasilinear.
For the existence result, its preferences need only
satisfy the monetary preference conditions.

\begin{proposition}[HIR and submodularity under quasilinear utilities]
\label{prop:quasiequivalence}
\normalfont
In the model of Section~\ref{subsec:itusetting},
suppose an ordinary agent $i$ has utility
\eqref{eq:quasilinear}. Then its preference satisfies
HIR if and only if its valuation $\mathrm{v}_i$
is submodular:
\begin{equation}\label{eq:quasiequivalence}
 \HIR
 \quad\Longleftrightarrow\quad
 \mathrm{v}_i\text{ is submodular}.
\end{equation}
The equivalence holds even when some portfolios have
value $-\infty$. Under either condition, the
technological domain
\[
 \mathcal D_i
 =
 \{\Psi\subseteq\Omega_i:
   \mathrm{v}_i(\Psi)>-\infty\}
\]
is closed under taking subsets.
\end{proposition}

For real-valued valuations in the quasilinear setting,
gross substitutability is stronger than submodularity
(\citealp{GS99,DFLP22}).
The submodularity condition in
Lemma~\ref{lem:submodular} applies to a utility
representation over bundles of contracts, which
already specify payments. In
Proposition~\ref{prop:quasiequivalence}, submodularity
instead applies to the valuation $\mathrm{v}_i$
over portfolios of agreements.

Quasilinear utilities satisfy the monetary preference
conditions. For each portfolio, utility is either
a continuous affine function of money or identically
$-\infty$. If $\mathrm{v}_i(\Psi)$ is finite,
finite willingness to pay holds with
\[
 \mathrm{b}_i(\Psi)
 =
 \mathrm{v}_i(\Psi)-\mathrm{v}_i(\varnothing).
\]
If $\mathrm{v}_i(\Psi)=-\infty$, the willingness-to-pay
condition holds vacuously.

Propositions~\ref{thm:main} and~\ref{prop:quasiequivalence}
therefore imply that a setwise stable outcome exists
when each ordinary agent's valuation is submodular
and the regulator satisfies the monetary preference
conditions.

\section{Deletion alignment}\label{sec:job}

\subsection{A generalization}

The selection principle extends beyond HIR when the regulator
supports ordinary agents' profitable contract deletions.
Consider the framework of Section~\ref{sec:framework}
with a finite contract set $X$, so that an outcome is any subset
of $X$. Each contract involves the regulator and one or more
ordinary agents. Preferences may include indifferences.

\begin{definition}[Deletion alignment]\label{def:deletion-alignment}
\normalfont
The regulator's preference satisfies \textbf{deletion alignment} (DEL)
with respect to ordinary agent $i\in J$ if, for every outcome
$Y\subseteq X$ and every $D\subseteq Y_i$,
\begin{equation}\label{eq:deletion-alignment}\tag{DEL}
 Y_i\setminus D\succ_i Y_i
 \quad\Longrightarrow\quad
 Y\setminus D\succ_r Y.
\end{equation}
\end{definition}

Deletion alignment requires the regulator to support every contract
deletion that strictly benefits the agent. Under DEL with respect
to $i$, any outcome that is individually rational for the regulator
is also individually rational for $i$. Otherwise, a deletion that
strictly benefits $i$ would also strictly benefit the regulator. 

\begin{proposition}[Matching under deletion alignment]
\label{prop:job-del}
\normalfont
In the finite-contract market, suppose that, for each
ordinary agent $i\in J$, either $i$ satisfies HIR or the regulator's
preference satisfies DEL with respect to $i$. Then a setwise stable
outcome exists. Moreover, every individually rational outcome that
the regulator weakly prefers to all other individually rational
outcomes is setwise stable.
\end{proposition}

Since $\mathcal A$ is finite and contains $\varnothing$, choose
$Y^r\in\mathcal A$ that the regulator weakly prefers to every other
outcome in $\mathcal A$. Suppose $Y^r$ is setwise blocked by $Z$
with final outcome $Y^*$. The participants, including the regulator,
receive individually rational bundles. Outsiders can only lose
contracts, so HIR ensures individual rationality for outsiders
satisfying HIR. For each remaining outsider, DEL and the regulator's
individual rationality imply that the outsider's final bundle is
individually rational. Thus $Y^*\in\mathcal A$, but the regulator
strictly prefers $Y^*$ to $Y^r$, contradicting the choice of $Y^r$.

If every ordinary agent satisfies HIR, this is the finite-contract
existence result obtained from Proposition~\ref{lem:selection}, and
the regulator's preference is unrestricted. Alternatively, ordinary
agents' preferences may be arbitrary if the regulator satisfies DEL
with respect to every ordinary agent. The proposition also allows
some agents to satisfy HIR and others to be covered by DEL.

\begin{remark}[Continuous transfers]
In the framework of Section~\ref{sec:tu}, define DEL by
\eqref{eq:deletion-alignment} for every outcome $Y$ and every
$D\subseteq Y_i$. The same argument shows that, if each ordinary
agent either satisfies HIR or is covered by DEL, any individually
rational outcome that the regulator weakly prefers to all other
individually rational outcomes is setwise stable. Thus an analogous
existence result holds whenever the regulator's optimum over
$\mathcal A$ is attained. Attainment requires a separate argument:
Lemma~\ref{lem:compact} assumes HIR for every ordinary agent and
cannot be invoked directly when some agents fail HIR but are covered
by DEL.
\end{remark}

\subsection{Application: project staffing within a corporate group}
\label{subsec:job-application}

Consider a corporate group whose headquarters approves staffing assignments across its project units. Each unit manages a project and selects a team of workers, while each worker accepts at most one assignment contract. Headquarters participates in every assignment agreement and has authority to approve or reject it. Let $F$ and $W$ be finite, disjoint sets of project units and workers, with $J=F\cup W$ and $r\notin F\cup W$. The regulator $r$ represents headquarters. A contract may assign one worker or a group of workers jointly to a single project unit. Its participants are therefore

\[
 \mathrm{N}(x)=\{r,f\}\cup T,
 \qquad f\in F,\quad \varnothing\ne T\subseteq W.
\]

Each contract specifies the project unit, the participating workers, their duties, and fixed remuneration for a given assignment
period. It requires the unit's and all participating workers' consent
as well as headquarters' approval. 
In the classical many-to-one job matching, every contract involves a single worker. We allow a contract to involve several workers, such as jointly assigning an engineering team to a project. In this collective-assignment setting, the identities and duties of the other workers are part of the contract and may affect a worker's ranking of that contract.

Workers have unit-demand preferences: they rank every bundle containing two or more assignment contracts below the empty bundle.
Their rankings of singleton bundles and unemployment are otherwise
unrestricted and may include indifferences. Thus
\[
 \mathcal A_w
 =\{\varnothing\}\cup
   \bigl\{\{x\}:x\in X_w,\ \{x\}\succeq_w\varnothing\bigr\},
 \qquad w\in W,
\]
so every worker satisfies HIR. 

Let $\mathrm{u}_f(A)$ denote unit $f$'s net project return from bundle $A$, after accounting for labor costs and other
contractual obligations. We assume that the unit's manager evaluates staffing arrangements according to this return. Complementarities
may arise across contracts: for example, a project may require both a design team and an installation team, assigned under separate
contracts, to become operational. The units' preferences need not
satisfy HIR, and their net returns need not be submodular.

Headquarters evaluates staffing arrangements using a
positively weighted sum of the units' net returns. The weights may reflect the relative importance of the projects. This objective aligns headquarters' evaluation of contract deletions with that of each unit and yields DEL.

\begin{proposition}[Centralized project staffing]
\label{prop:job-weighted}
\normalfont
In the finite-contract staffing market described above,
let $\mathrm{u}_f:2^{X_f}\to\R$ represent project unit $f$'s preference
over its assignment bundles. If headquarters' preference is
represented by
\begin{equation}\label{eq:job-regulator-objective}
 \mathrm{u}_r(Y)
 =\sum_{f\in F}\lambda_f\mathrm{u}_f(Y_f),
 \qquad \lambda_f>0,
\end{equation}
then it satisfies DEL with respect to every unit.
Consequently, the finite-contract market admits a setwise stable
outcome for arbitrary preferences of the project units and arbitrary
unit-demand preferences of the workers.
\end{proposition}

Each assignment contract involves exactly one project unit,
even when it involves several workers. For any $D\subseteq Y_f$,
deleting $D$ therefore leaves every other unit's bundle unchanged, so
\[
 \mathrm{u}_r(Y\setminus D)-\mathrm{u}_r(Y)
 =\lambda_f\bigl[
   \mathrm{u}_f(Y_f\setminus D)-\mathrm{u}_f(Y_f)
  \bigr].
\]
Since $\lambda_f>0$, any deletion that strictly increases
the unit's net return also strictly increases headquarters' utility.
Thus headquarters satisfies DEL with respect to every
project unit. Since every worker satisfies HIR,
Proposition~\ref{prop:job-del} guarantees the existence of a setwise
stable outcome.

The objective need not be additive: any function of the
units' returns that is strictly increasing in each coordinate also
induces a preference satisfying DEL. Headquarters may also incur
central administration costs. The same conclusion holds if its
objective is
\begin{equation*}
 \mathrm{u}_r(Y)
 =\sum_{f\in F}\lambda_f\mathrm{u}_f(Y_f)-\mathrm{K}(Y),
\end{equation*}
where $\mathrm{K}:2^X\to\R_+$ is an administration cost
function that does not increase when contracts are removed. Such
costs may include maintaining personnel records, processing payments,
and coordinating assignments.

\section{Appendix}

\subsection{Proof of Lemma~\ref{lem:compact}}

Fix a set of executed agreements $\Psi\subseteq\Omega$.
Consider any individually rational outcome $Y$ with
$\mathrm{\uptau}(Y)=\Psi$.
For each contract $(w,\mathbf t^w)\in Y$ and ordinary
signatory $i\in\mathrm{N}(w)\setminus\{r\}$, HIR implies
that the singleton bundle
$\{(w,\mathbf t^w)\}\subseteq Y_i$
is individually rational.
Comparing this bundle with the empty bundle and
applying \eqref{eq:fwtp} gives
\begin{equation}\label{eq:ordinaryupper}
 t_i^w\le\mathrm{b}_i(\{w\})=:\mathrm{b}_i^w.
\end{equation}
Transfer balance therefore gives a lower bound on
the regulator's payment:
\begin{equation}\label{eq:reglower}
 t_r^w
 =-\sum_{i\in\mathrm{N}(w)\setminus\{r\}}t_i^w
 \ge
 -\sum_{i\in\mathrm{N}(w)\setminus\{r\}}
   \mathrm{b}_i^w
 =:L^w.
\end{equation}
The regulator's individual rationality and
\eqref{eq:fwtp} imply
\[
 \sum_{w\in\Psi}t_r^w\le\mathrm{b}_r(\Psi).
\]
Hence, for each $w\in\Psi$,
\begin{equation}\label{eq:regupper}
 t_r^w
 \le
 \mathrm{b}_r(\Psi)
 -\sum_{v\in\Psi\setminus\{w\}}L^v
 =:\mathrm{H}^w(\Psi).
\end{equation}
For an ordinary signatory $i$ of $w$, transfer balance
once again implies
\begin{equation}\label{eq:ordinarylower}
 t_i^w
 =-t_r^w
  -\sum_{j\in\mathrm{N}(w)\setminus\{r,i\}}t_j^w
 \ge
 -\mathrm{H}^w(\Psi)
 -\sum_{j\in\mathrm{N}(w)\setminus\{r,i\}}
   \mathrm{b}_j^w.
\end{equation}
These inequalities provide finite upper and lower
bounds for every transfer coordinate.
The bounds depend only on the fixed agreement set
$\Psi$ and the willingness-to-pay bounds.

To establish closedness without subtracting
extended-real utilities, write
$\mathbf t=(\mathbf t^w)_{w\in\Psi}$
for the transfer coordinates.
For each agent $i$ and each fixed portfolio
$\Phi\subseteq\Psi_i$, define
\[
 \mathrm{W}_{i,\Phi}(\mathbf t)
 =
 \exp\left(
 \mathrm{V}_i\left(
 \Phi,-\sum_{w\in\Phi}t_i^w
 \right)
 \right),
 \qquad
 \exp(-\infty)=0.
\]
Each $\mathrm{W}_{i,\Phi}$ is a continuous real-valued
function of the transfer coordinates.
The portfolios $\Phi\subseteq\Psi_i$ index all
subbundles of agent $i$'s bundle.
Individual rationality is therefore equivalent to
the finitely many closed inequalities
\[
 \mathrm{W}_{i,\Psi_i}(\mathbf t)
 \ge
 \mathrm{W}_{i,\Phi}(\mathbf t)
 \qquad
 (i\in I,\ \Phi\subseteq\Psi_i).
\]
The comparison with $\Phi=\varnothing$ ensures that
\[
 \mathrm{W}_{i,\Psi_i}(\mathbf t)
 \ge
 \exp\bigl(\mathrm{V}_i(\varnothing,0)\bigr)>0,
\]
including at every limit point.
Transfer balance imposes closed linear equalities.
The set of transfers implementing an individually
rational outcome is thus a closed and bounded subset
of a finite-dimensional Euclidean space, and hence
is compact.
When $\Psi=\varnothing$, the transfer set consists
of the unique vector with no coordinates and is
also compact.

\subsection{Proof of Proposition~\ref{prop:gshir}}

For part~(i), fix
$\Psi'\subseteq\Psi\subseteq\Omega_i$,
a payment vector $\mathbf p_i\in\R^{\Omega_i}$,
and $\Phi\in\mathcal C_i(\Psi;\mathbf p_i)$.
Keep the payments on $\Psi$ fixed and choose
sufficiently large positive payments outside $\Psi$,
obtaining a vector $\widehat{\mathbf p}_i$ such that
\[
 \sum_{w\in\Gamma}\widehat p_i^w
 >
 \mathrm{b}_i(\Gamma)
 \qquad
 \text{for every }\Gamma\subseteq\Omega_i
 \text{ with }\Gamma\not\subseteq\Psi.
\]
Such a choice is possible because there are finitely
many portfolios, every $\Gamma\not\subseteq\Psi$
contains at least one agreement outside $\Psi$,
and the payments on $\Psi$ remain fixed.
By \eqref{eq:fwtp},
\[
 \mathrm{V}_i\left(
 \Gamma,-\sum_{w\in\Gamma}\widehat p_i^w
 \right)
 <
 \mathrm{V}_i(\varnothing,0)
 \qquad
 \text{whenever }\Gamma\not\subseteq\Psi.
\]
Thus no portfolio containing an agreement outside
$\Psi$ is demanded, and
\[
 \mathrm{D}_i(\widehat{\mathbf p}_i)
 =
 \mathcal C_i(\Psi;\mathbf p_i).
\]

Next, increase the payments outside $\Psi'$
sufficiently, keeping those on $\Psi'$ unchanged.
The same argument yields a vector
$\mathbf q_i\ge\widehat{\mathbf p}_i$ such that
\[
 \mathrm{D}_i(\mathbf q_i)
 =
 \mathcal C_i(\Psi';\mathbf p_i).
\]
Since $\Phi\in\mathrm{D}_i(\widehat{\mathbf p}_i)$
and every payment on $\Psi'$ is unchanged,
GS yields a demanded portfolio
$\Phi'\in\mathrm{D}_i(\mathbf q_i)$ containing
$\Phi\cap\Psi'$.
This establishes monetary admission monotonicity.

For part~(ii), let $S$ be an individually rational
bundle of contracts and write
$\Phi=\mathrm{\uptau}(S)$.
Let $\mathbf p_i$ record the agent's payments in $S$,
with arbitrary entries outside $\Phi$.
Then
$\Phi\in\mathcal C_i(\Phi;\mathbf p_i)$.
For any $\Psi'\subseteq\Phi$, apply
\eqref{eq:gssone} with $\Psi=\Phi$.
It yields
$\Phi'\in\mathcal C_i(\Psi';\mathbf p_i)$
such that
\[
 \Psi'=\Phi\cap\Psi'
 \subseteq\Phi'
 \subseteq\Psi'.
\]
Hence $\Phi'=\Psi'$, so the corresponding subbundle
of $S$ is individually rational.
Since $\Psi'$ is arbitrary, HIR follows.

\subsection{Proof of Proposition~\ref{prop:procurement-income-hir}}

The outside option is
$\mathrm{V}_i(\varnothing,0)=0$.
Continuity follows from the assumptions on
$\mathrm{h}_i$.
If $\mathrm{V}_i(\Psi,-P)\ge0$, then
\[
 \mathrm{h}_i\bigl(\mathrm{v}_i(\Psi)-P\bigr)
 \ge
 \sum_{w\in\Psi}c_{iw}
 \ge0.
\]
Since $\mathrm{h}_i$ is strictly increasing and
$\mathrm{h}_i(0)=0$, we have
$P\le\mathrm{v}_i(\Psi)$.
Thus finite willingness to pay holds with
$\mathrm{b}_i(\Psi)=\mathrm{v}_i(\Psi)$.

To prove HIR, fix an individually rational bundle
$A\in\mathcal A_i$.
For every $B\subseteq A$, define its net financial
contribution by
\begin{equation}\label{eq:procurement-cash}
 \mathrm{M}_i(B)
 =
 \mathrm{v}_i\bigl(\mathrm{\uptau}(B)\bigr)
 -
 \sum_{(w,\mathbf t^w)\in B}t_i^w.
\end{equation}
Since each agreement appears at most once in $A$,
the agreement map preserves unions and intersections
of subsets of $A$.
Submodularity of $\mathrm{v}_i$ and additivity of
payments therefore imply that $\mathrm{M}_i$ is
submodular on $2^A$.

For $x\in B\subseteq A$, write
\[
 \mathrm{d}_i(x;B)
 =
 \mathrm{M}_i(B)
 -
 \mathrm{M}_i(B\setminus\{x\}).
\]
Individual rationality of $A$ gives, for each
$x=(w,\mathbf t^w)\in A$,
\[
 \mathrm{h}_i(\mathrm{M}_i(A))
 -
 \mathrm{h}_i\bigl(
 \mathrm{M}_i(A)-\mathrm{d}_i(x;A)
 \bigr)
 \ge c_{iw}\ge0.
\]
If $\mathrm{d}_i(x;A)<0$, strict monotonicity of
$\mathrm{h}_i$ would make the left-hand side negative.
Hence $\mathrm{d}_i(x;A)\ge0$.
Applying submodularity to $B$ and
$A\setminus\{x\}$ yields
\begin{equation}\label{eq:procurement-diminishing}
 \mathrm{d}_i(x;B)
 \ge
 \mathrm{d}_i(x;A)
 \ge0
 \qquad
 \text{for every }x\in B\subseteq A.
\end{equation}

We next show that
$\mathrm{M}_i(B)\le\mathrm{M}_i(A)$
for every $B\subseteq A$.
The conclusion is immediate if $B=A$.
Otherwise, write
$A\setminus B=\{x_1,\ldots,x_k\}$ and add these
contracts to $B$ one at a time.
Let
\[
 T_0=B,\qquad
 T_j=B\cup\{x_1,\ldots,x_j\}
 \quad(j=1,\ldots,k),
 \qquad T_k=A.
\]
For each $j$, we have $x_j\in T_j\subseteq A$ and
$T_j\setminus\{x_j\}=T_{j-1}$.
Applying \eqref{eq:procurement-diminishing} with
$x=x_j$ and $B=T_j$ gives
\[
 \mathrm{M}_i(T_j)-\mathrm{M}_i(T_{j-1})
 =
 \mathrm{d}_i(x_j;T_j)
 \ge
 \mathrm{d}_i(x_j;A)
 \ge0.
\]
Summing over $j=1,\ldots,k$ and cancelling the
intermediate terms yields
\[
 \mathrm{M}_i(A)-\mathrm{M}_i(B)
 =
 \sum_{j=1}^k
 \bigl[
 \mathrm{M}_i(T_j)-\mathrm{M}_i(T_{j-1})
 \bigr]
 \ge
 \sum_{j=1}^k\mathrm{d}_i(x_j;A)
 \ge0.
\]

For fixed $d\ge0$, concavity of $\mathrm{h}_i$
implies that
$\mathrm{h}_i(m)-\mathrm{h}_i(m-d)$
is nonincreasing in $m$.
For fixed $m$, monotonicity of $\mathrm{h}_i$
implies that this increment is nondecreasing in $d$.
Combining these observations with
$\mathrm{M}_i(B)\le\mathrm{M}_i(A)$ and
\eqref{eq:procurement-diminishing}, we obtain,
for $x=(w,\mathbf t^w)\in B\subseteq A$,
\begin{align}
 \mathrm{u}_i(B)-\mathrm{u}_i(B\setminus\{x\})
 &=
 \mathrm{h}_i(\mathrm{M}_i(B))
 -
 \mathrm{h}_i\bigl(
 \mathrm{M}_i(B)-\mathrm{d}_i(x;B)
 \bigr)
 -c_{iw}
 \notag\\
 &\ge
 \mathrm{h}_i(\mathrm{M}_i(A))
 -
 \mathrm{h}_i\bigl(
 \mathrm{M}_i(A)-\mathrm{d}_i(x;A)
 \bigr)
 -c_{iw}
 \notag\\
 &=
 \mathrm{u}_i(A)-\mathrm{u}_i(A\setminus\{x\})
 \ge0.
 \label{eq:procurement-marginal}
\end{align}
For any $D\subseteq B\subseteq A$, successively
delete the contracts in $B\setminus D$ from $B$.
Applying \eqref{eq:procurement-marginal} at each
step gives $\mathrm{u}_i(B)\ge\mathrm{u}_i(D)$.
Since $D$ is arbitrary, $B$ is individually rational.
This holds for every $B\subseteq A$, proving HIR.

Example~\ref{ex:procurement-income-nongs} shows that
this class includes preferences violating monetary
admission monotonicity.
By Proposition~\ref{prop:gshir}(i), these preferences
also violate GS.
The existence conclusion follows from
Proposition~\ref{thm:main}.

\subsection{Proof of Proposition~\ref{prop:quasiequivalence}}

First, suppose $\mathrm{v}_i$ is submodular.
The utility representation over contract bundles is
\[
 \mathrm{u}_i(B)
 =
 \mathrm{v}_i\bigl(\mathrm{\uptau}(B)\bigr)
 -
 \sum_{(w,\mathbf t^w)\in B}t_i^w,
 \qquad B\in\mathcal B_i.
\]
For any $B,D\in\mathcal B_i$ such that
$B\cup D\in\mathcal B_i$, the agreement map
preserves unions and intersections:
\[
 \begin{aligned}
 \mathrm{\uptau}(B\cup D)
 &=
 \mathrm{\uptau}(B)\cup\mathrm{\uptau}(D),\\
 \mathrm{\uptau}(B\cap D)
 &=
 \mathrm{\uptau}(B)\cap\mathrm{\uptau}(D).
 \end{aligned}
\]
The second equality holds because $B\cup D$
contains at most one contract for each agreement.
Submodularity of $\mathrm{v}_i$ and additivity of
payments therefore imply
\[
 \mathrm{u}_i(B)+\mathrm{u}_i(D)
 \ge
 \mathrm{u}_i(B\cup D)+\mathrm{u}_i(B\cap D).
\]
Thus $\mathrm{u}_i$ is submodular on $\mathcal B_i$.
Since
$\mathrm{u}_i(\varnothing)
 =\mathrm{v}_i(\varnothing)\in\R$,
Lemma~\ref{lem:submodular} implies HIR, including
when some portfolios have value $-\infty$.

Conversely, suppose agent $i$ satisfies HIR.
We suppress the subscript $i$ in the remainder
of the proof.\footnote{
We suppress the subscript on utilities, valuations,
payments, and the domain $\mathcal D_i$, but retain
$\Omega_i$ for the agent's set of agreements.}
For a fixed payment vector
$\mathbf p\in\R^{\Omega_i}$, define
\[
 \mathrm{F}^{\mathbf p}(\Psi)
 :=
 \mathrm{v}(\Psi)-\mathrm{p}(\Psi)
 =
 \mathrm{v}(\Psi)-\sum_{w\in\Psi}p^w,
 \qquad \Psi\subseteq\Omega_i.
\]
For any bundle $A\in\mathcal B_i$, write
$\Phi=\mathrm{\uptau}(A)$.
For each portfolio $\Psi\subseteq\Phi$, define
the corresponding subbundle by
\begin{equation}\label{eq:subbundle}
 A_\Psi
 =
 \{x\in A:\mathrm{\uptau}(x)\in\Psi\}.
\end{equation}
Let $\mathbf p$ record the agent's payments in
the contracts of $A$, with arbitrary entries
outside $\Phi$.
Then
\begin{equation}\label{eq:quasibundleutility}
 \mathrm{u}(A_\Psi)
 =
 \mathrm{v}(\Psi)-\sum_{w\in\Psi}p^w
 =
 \mathrm{F}^{\mathbf p}(\Psi)
 \qquad(\Psi\subseteq\Phi).
\end{equation}
This identity allows us to compare contract bundles
through their agreement portfolios.

Any specified personal payments on a portfolio can
be implemented by a bundle of contracts.
For each agreement $w$, assign payment $p^w$
to agent $i$, payment $-p^w$ to the regulator,
and zero to every other signatory.
The resulting bundles need not be individually
rational for the other signatories.

We first show that $\mathcal D$ is closed under
taking subsets.
Take a portfolio $\Phi$ with
$\mathrm{v}(\Phi)>-\infty$.
The case $\Phi=\varnothing$ is immediate.
For nonempty $\Phi$, construct a bundle $A$ with
$\mathrm{\uptau}(A)=\Phi$ and payment $-K$
in each contract.
For every proper subset $\Psi\subsetneq\Phi$
with $\mathrm{v}(\Psi)>-\infty$,
\[
 \mathrm{u}(A)-\mathrm{u}(A_\Psi)
 =
 \mathrm{v}(\Phi)-\mathrm{v}(\Psi)
 +
 K\bigl(|\Phi|-|\Psi|\bigr).
\]
There are finitely many such subsets, and the
coefficient of $K$ is positive in every comparison.
We can therefore choose a single $K>0$ large
enough to make all these differences strictly
positive.
Subbundles whose agreement portfolios have value
$-\infty$ also yield strictly lower utility.
Thus $A$ is individually rational.

By HIR, every subbundle $A_\Psi$ is individually
rational, so
\[
 \mathrm{u}(A_\Psi)
 \ge
 \mathrm{v}(\varnothing)>-\infty.
\]
Since its payments are finite,
$\mathrm{v}(\Psi)>-\infty$ for every
$\Psi\subseteq\Phi$.
This proves that $\mathcal D$ is closed under
taking subsets.

Now suppose, contrary to the desired conclusion,
that $\mathrm{v}$ is not submodular.
If the submodularity inequality fails for two
portfolios, their union must have finite value:
otherwise, the right-hand side of
\eqref{eq:submodularity} would be $-\infty$,
and the inequality could not fail.
Denote this union by $U\subseteq\Omega_i$.
The preceding argument implies that
$\mathrm{v}$ is real-valued on $2^U$.

The elementary two-agreement characterization of
submodularity therefore yields a portfolio
$\Psi\subseteq U$ and distinct agreements
$w,v\in U\setminus\Psi$ such that
\begin{equation}\label{eq:positivecomplementarity}
 \Delta
 =
 \mathrm{v}(\Psi\cup\{w,v\})
 +\mathrm{v}(\Psi)
 -\mathrm{v}(\Psi\cup\{w\})
 -\mathrm{v}(\Psi\cup\{v\})
 >0.
\end{equation}
All four values are finite.
This characterization follows by iterating the
elementary inequalities to obtain diminishing
marginal values on nested sets, then summing the
marginal comparisons to recover the submodularity
inequality.

Choose $0<\varepsilon<\Delta/2$ and set
\[
 \begin{aligned}
 p^w
 &=
 \mathrm{v}(\Psi\cup\{w\})
 -\mathrm{v}(\Psi)+\varepsilon,\\
 p^v
 &=
 \mathrm{v}(\Psi\cup\{v\})
 -\mathrm{v}(\Psi)+\varepsilon.
 \end{aligned}
\]
Assign $p^\omega=-K$ for $\omega\in\Psi$,
write $\Phi=\Psi\cup\{w,v\}$, and construct
a bundle $A$ with agreement portfolio $\Phi$
and these personal payments.
By \eqref{eq:quasibundleutility},
\begin{align*}
 \mathrm{F}^{\mathbf p}(\Psi\cup\{w\})
 -
 \mathrm{F}^{\mathbf p}(\Psi)
 &=-\varepsilon,\\
 \mathrm{F}^{\mathbf p}(\Psi\cup\{v\})
 -
 \mathrm{F}^{\mathbf p}(\Psi)
 &=-\varepsilon,\\
 \mathrm{F}^{\mathbf p}(\Phi)
 -
 \mathrm{F}^{\mathbf p}(\Psi)
 &=\Delta-2\varepsilon>0.
\end{align*}
These comparisons show that $A$ is strictly preferred
to every proper subbundle retaining all contracts
for agreements in $\Psi$.

For a portfolio $\Gamma\subseteq\Phi$ that omits
at least one agreement in $\Psi$,
\[
 \begin{aligned}
 \mathrm{F}^{\mathbf p}(\Phi)
 -
 \mathrm{F}^{\mathbf p}(\Gamma)
 &=
 \mathrm{v}(\Phi)-\mathrm{v}(\Gamma)
 +K|\Psi\setminus\Gamma|\\
 &\quad
 -\sum_{\omega\in\{w,v\}\setminus\Gamma}p^\omega.
 \end{aligned}
\]
All valuation terms are finite, the coefficient
$|\Psi\setminus\Gamma|$ is positive, and the
remaining terms do not depend on $K$.
Since there are finitely many such portfolios,
a single sufficiently large $K>0$ makes all
these differences positive.
If $\Psi=\varnothing$, no additional comparisons
are needed.
Hence $A$ is individually rational.

However, the subbundle $A_{\Psi\cup\{w\}}$ is
not individually rational: deleting the contract
for agreement $w$ leaves $A_\Psi$ and raises
utility by $\varepsilon$.
This contradicts HIR and proves submodularity.

\end{document}